<

# Analysis and Design of Double-Transmitting Coil Systems based on Parity-Time Symmetry

Xiaokui Kang, Hongbin Ma, Zihui Liu, Yuanyuan Wang, Jiangtao Huangfu, Rui Xi, Ying Li

***Abstract*—The wireless power transfer (WPT) system based on parity-time (PT) symmetry has the advantages of robustness, stability, and efficient power transmission. However, traditional PT symmetry structures have limited voltage power output and are susceptible to horizontal misalignment effects. Multiple transmission coils have been proven to improve the power and misalignment tolerance of WPT systems, but variations in inter-coil coupling significantly affect transmission power. Towards this end, this article proposes a double-transmitting coil WPT system based on PT symmetry, which is innovative in that a novel negative resistance structure based on operational amplifier (OA) is proposed, and the parallel structure of double-transmitting coil is applied to this negative resistance structure. Compared with the traditional WPT system based on a single-transmitting coil PT, this system improves the power and misalignment tolerance in the PT symmetry region. The large-sized and small-sized receiving coils increase the load power to 313% and 185% respectively. Moreover, in the symmetry region, the displacement of the receiving coil in the horizontal and vertical directions results in variations of the equivalent coupling coefficient, which can achieve stable power transmission in three dimensions. The voltage fluctuation rate of both receiving coils does not exceed 3.4%, greatly improving the degree of freedom of the receiving system.**

***Index Terms*—parity-time (PT) symmetry, wireless power transfer (WPT), double-transmitting coil, power improvement.**

Manuscript received date; revised date; accepted date. Date of publication date; date of current version date. This work was supported by the National Key Research and Development Program of China under Grant No. 2022YFA1405201, the Zhejiang Provincial Natural Science Foundation of China under Grant No.LZ24A050002 and No.LR26A050001, the Fundamental Research Funds for the Central Universities, and the Innovation Fund of Xidian University under Grant No. XJSJ24094. (Corresponding author: Jiangtao Huangfu, Rui Xi, Ying Li.)

Ying Li is with the State Key Laboratory of Extreme Photonics and Instrumentation, Zhejiang Key Laboratory of Intelligent Electromagnetic Control and Advanced Electronic Integration, Zhejiang University, Hangzhou 310027, China (e-mail: eleying@zju.edu.cn).

Ying Li, Jiangtao Huangfu and Hongbin Ma are with the College of Information Science and Electronic Engineering, Zhejiang University, Hangzhou 310027, China (e-mail: huangfujt@zju.edu.cn).

Xiaokui Kang, Zihui Liu, Yuanyuan Wang and Rui Xi are with the Hangzhou Institute of Technology, Xidian University, Hangzhou 311231, China(e-mail: xirui@xidian.edu.cn).

Xiaokui Kang and Hongbin Ma contributed equally to this work.

## I. INTRODUCTION

WIRELESS power transfer (WPT) technology offers advantages such as electrical isolation, safety, and convenience, demonstrating significant application potential in consumer electronics like drones, electric vehicles, and implantable biomedical devices in recent years [1], [2]. Among various WPT technologies, magnetic resonant WPT has stood out due to its unique advantages, sparking a research boom across a wide range of applications. Stable output power is one of the key application goals pursued by WPT technology. However, magnetic resonance lacks robustness when mutual inductance changes, leading to fluctuations in transmission power and efficiency. To achieve robust and efficient power transfer under varying mutual inductance conditions, methods such as parameter estimation [3], [4], frequency tracking [5], [6], impedance matching [7], [8], and coupling coil optimization [9], [10], [11] can be employed. In [12], which introduced topological metamaterials into magnetically coupled coils to enhance energy transfer efficiency to a certain extent. However, the structure is complex and requires a relatively large number of metamaterials. Therefore, most of these methods inevitably suffer from drawbacks like complex control and low transmission efficiency, and system feedback control also relies on wireless communication between the transmitter and the receiver.

In 2017, the reference [13] introduced the parity-time (PT) symmetry into WPT, providing an interdisciplinary solution to enhance the robustness of WPT systems. Within the precise PT symmetry region, PT based on WPT can automatically select the operating frequency corresponding to the highest efficiency, ensuring robust power transmission across a wide range of coupling coefficients without requiring any active tuning or feedback control. For this WPT system, a negative resistance circuit composed of operational amplifier (OA) is employed to provide gain. The reference [14] proposed a novel wireless power supply structure to form a microwave generator, where the entire system including a transmitter, a receiver and an intermediate space. An in-depth theoretical study of self-excited oscillation systems is conducted in [15], proposing stability criteria and initiation criteria. Traditionally, experimental validation was mainly performed on two distinct topologies: series-series (SS) and parallel-parallel (PP). The SS topology achieved a transmission distance of 32 cm and an output power of 2.5 W, while the PP topology only reached 23 cm and 70 mW. A nonlinear PT symmetry WPT system based on SS structure was proposed in [16], which improved the gain of the circuit by connecting two amplifiers negative resistance circuits in series, thereby achieving high power, high efficiency, and a wider energy output range. In [17][18], the OA based on PT symmetry circuit proposed in the paper increased the transmission distance significantly while ensuring stable power transmission by adding intermediate coil between the transmitting coil and the receiving coil. However, equal coupling coefficients must be satisfied between adjacent resonant cavities, which requires precise mechanical control and is difficult to achieve in practice. The article [19] explored a robust maximum power point tracking PT symmetry WPT system with implementing series circuits. The article [20] proposed PP circuit with the use of planar or cylindrical coils as transmitters and receivers to achieve PT symmetry and ensure stable power transfer, releasing the restriction that the receiving coil and transmitting coil must remain consistent. Reference [21] addressed these limitations by adopting anti-PT symmetry. By adding a bypass capacitor to create a synthetic dimension, greatly broadening the transmission range. In [22] , which investigated the effect of the level pinning phenomenon in anti-PT symmetry and explored the influence of the transmitter/receiver coil size ratio on wireless power transfer.

Subsequently, gain element implementation strategies based on bridge inverters and E-class inverters were proposed in [23], [24], and [25], which greatly improved the transmission efficiency and power of PT based on WPT systems. However, the negative resistance gain circuit composed of bridge inverters and E-class inverters would make the circuit complex and require the construction of external control circuits and input control signals to ensure normal operation, greatly increasing cost and complexity. The robustness of PT based on WPT systems in strongly coupled regions is limited by transmission distance. In [26], compensating inductance is connected in series with the receiving coil to enhance the transmission distance, but this structure reduces transmission efficiency and increases the size and complexity of the receiving circuit. A high-order PT symmetry system using S-PS compensation was proposed in [27], which parallel compensating capacitors to the receiving circuit for the equivalent load resistance reduction, which improved transmission distance but also reduce transmission efficiency and increase complexity. A WPT system based on a combination of electrical coupling and magnetic coupling was proposed in [28], which greatly increased transmission distance and power, but significantly increases cost and circuit complexity. A two-to-one power transfer circuit based on SS circuit was proposed in [29], which can transfer power to different receiving circuits by changing the compensating capacitor at the transmitting end. The reference [30] presented an excellent high-power (30W) solution using inverters and SS topology, targeting applications like drone charging. However, the inverter architecture requires bulky gate drivers and external control loops, rendering it unsuitable for space constrained, milli watt level applications such as biomedical implants or on chip power transfer.

In summary, current PT-WPT systems based on OA are limited to one-to-one configurations, exhibiting low output power and weak spatial misalignment tolerance. Power enhancement typically relies on increasing the number of amplifiers, which leads to inefficient resource utilization, higher circuit complexity, and constrained application scenarios. Meanwhile, most existing research on inverters based on SS circuit as gain elements is only applicable to high-power, large-scale applications such as drones or industrial robots. Their complex external drive signal structures are clearly unsuitable for miniaturized devices. Furthermore, the SS circuit topology is primarily effective in low-load conditions. In contrast, the OA architecture facilitates integrated miniaturization—requiring only power supply to the amplifier to achieve self-oscillating energy transmission making it highly suitable for small scale applications such as implantable medical devices and capsule robots.

The main work and contributions are summarized as following:

1) Based on the OA negative resistance self-oscillation structure, a WPT system with dual-transmitting and single receiving coils is designed. The circuit adopts a PP structure and can operate stably under high impedance load conditions.

2) An approximate circuit simplified model of the proposed WPT system was established, and the operating conditions and transmission characteristics of two different sized receiving coil systems in the PT symmetry region were analyzed. Moreover, the feasibility of the approximate conditions has been experimentally verified.

3) Both types of receiving coil structures can increase the load voltage, and the size of the receiving coil will affect the voltage at the receiving end. The size of the receiving coil can be selected according to the actual application scenario.

4) Due to the dual transmitting coils, the magnetic field coupling range is increased, which enables the position movement of the receiving coil in three-dimensional directions and increases the positional misalignment freedom of the receiving system.

The rest of this article is organized as follows. In the second section, the circuit model of the proposed system was established, and the equivalent circuit under two different sizes of receiving coils was analyzed. The operating frequency and critical coupling conditions were obtained, and compared with the traditional single-transmitting coil PT WPT system. In the third section, the performance parameters of two types of receiving coils were compared and analyzed. In the fourth section, the hardware experimental scheme was introduced, and an experimental prototype based on an OA was presented. Finally, the fifth section provides a summary of this article.

## II. THEORETICAL ANALYSIS AND MODELING

### *A. System Modeling and Analysis*

There are four common topologies for WPT based on PT symmetry: SS, PS, PP, and SP, among which only SS and PP structures can achieve stable transmission. This article focuses on analyzing the PP structure. WPT structures usually consist of OA with negative resistors to provide energy to the circuit.

The equivalent circuit diagram of one-to-one WPT system is shown in Figure 1, where $R_{p1}$ and $L_1$ are the internal resistance

<

**Fig. 1.** WPT equivalent circuit diagram (one-to-one).

and self-inductance of the transmitting coil, $R_{p2}$ and $L_2$ are the internal resistance and self-inductance of the receiving coil, $C_1$ and $C_2$ are compensation capacitors, $M$ is mutual inductance, $R_L$ is load resistance, and is the equivalent negative resistance of the OA circuit. Using circuit principles to model and analyze the system, ignoring the influence of coil internal resistance, and utilizing Kirchhoff's voltage and current laws, we can obtain:

$$\begin{aligned} U_i &= j\omega L_1 I_1 + j\omega M I_2 \\ U_L &= j\omega L_2 I_2 + j\omega M I_1 \\ I_1 + j\omega C_1 U_i - \frac{U_i}{R_N} &= 0 \\ I_2 + j\omega C_2 U_L + \frac{U_L}{R_L} &= 0 \end{aligned} \quad (1)$$

Among them, $U_{\mathrm{i}}$ and $U_L$ represent the voltages across the read coil and sensing coil, respectively; $I_1$ and $I_2$ are the currents flowing through the transmitting coil and receiving coil, where $M$ is the mutual inductance between the transmitting and receiving coils. The equation is transformed into a matrix form expressed in terms of $U_{\mathrm{i}}$ and $U_L$ as follows:

$$\begin{bmatrix} 1-\omega^2 L_1 C_1 - j\dfrac{\omega L_1}{R_N} & -\omega^2 M C_2 + j\dfrac{\omega M}{R_L} \\ -\omega^2 M C_1 - j\dfrac{\omega M}{R_N} & 1-\omega^2 L_2 C_2 + j\dfrac{\omega L_2}{R_L} \end{bmatrix} \begin{bmatrix} U_i \\ U_L \end{bmatrix} = 0 \quad (2)$$

$L_1 = L_2 = L$ , $C_1 = C_2 = C$ , $R_N = R_L$ . Under the combined transformation of spatial inversion and temporal inversion , the system equation remains unchanged, indicating that the system satisfies PT symmetry. To make a matrix equation have a solution, the determinant of its coefficient matrix must be 0, and the $\omega$ resulting expression is:

$$\omega_{1,2} = \sqrt{\frac{LC - \dfrac{L^2 - M^2}{2R_L^{\,2}} \pm \sqrt{\left(LC - \dfrac{L^2 - M^2}{2R_L^{\,2}}\right)^2 - C^2(L^2 - M^2)}}{C^2(L^2 - M^2)}} \quad (3)$$

The quality factor of the receiving end is $Q = R_{\mathrm{L}}(C/L)^{1/2}$ , and the expression for is:

$$\omega_{1,2} = \sqrt{\frac{k^2 + 2Q^2 - 1 \pm \sqrt{1 - 4Q^2 + (4Q^4 + 4Q^2 - 2)k^2 + k^4}}{2Q^2(1-k^2)LC}} \quad (4)$$

To ensure $\omega$ is a real number, the expression for the critical coupling coefficient $k_C$ can be obtained:

$$k_C = \sqrt{1 - 2Q^2 - 2Q^4 + 2Q^3\sqrt{Q^2 + 2}} \quad (5)$$

The system satisfies PT symmetry, and at this point: $U_L / U_{\mathrm{i}} = 1$ .The theoretical efficiency can be obtained as follows:

$$\eta = \frac{P_L}{P_{\mathrm{i}}} = \frac{U_{\mathrm{L}}{}^2 R_N}{U_{\mathrm{i}}^2 R_L} = 100\% \quad (6)$$

In one configuration, the transmitting and receiving coils are of equal size. This is designated as system A, and the receiving coil A can only be fully coupled with one of them at the same time. As shown in Figure 2. Let $M_{12}$ and $M_{32}$ be the coupling coefficients of the double-transmitting coil and the receiving coil, respectively.

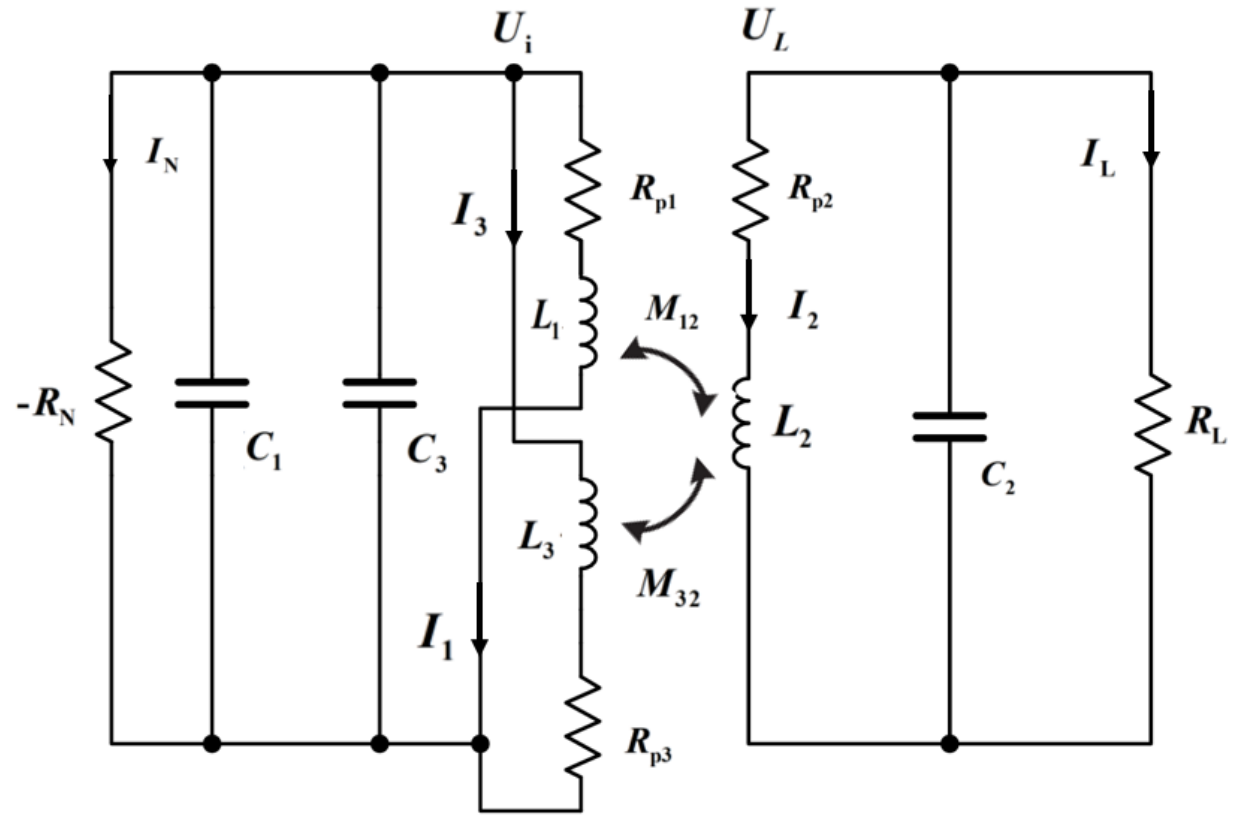

**Fig. 2.** Equivalent circuit diagram of the two-to-one system A

$$\begin{aligned} -R_{\mathrm{N}} I_{\mathrm{N}} &= j\omega L_1 I_1 + j\omega M_{12} I_2 \\ -R_{\mathrm{N}} I_{\mathrm{N}} &= j\omega L_3 I_3 + j\omega M_{32} I_2 \\ R_{\mathrm{L}} I_{\mathrm{L}} &= j\omega M_{12} I_1 + j\omega M_{32} I_3 + j\omega L_2 I_2 \\ I_1 + I_3 - j\omega(C_1 + C_3) R_{\mathrm{N}} I_{\mathrm{N}} + I_{\mathrm{N}} &= 0 \\ I_2 + j\omega C_2 R_{\mathrm{L}} I_{\mathrm{L}} + I_{\mathrm{L}} &= 0 \end{aligned} \quad (7)$$

The matrix forms include:

$$\begin{bmatrix} \dfrac{\omega}{\omega_T} + j\gamma_T - \dfrac{\omega_T}{\omega}\dfrac{1}{(1-k_{TR}^2)} & -\dfrac{\omega_R}{\omega}\dfrac{k_{PS}}{(1-k_{TR}^2)}\dfrac{\gamma_T}{\gamma_R}\sqrt{\dfrac{L_R}{L_T}} \\ -\dfrac{\omega_T}{\omega}\dfrac{k_{PS}}{(1-k_{TR}^2)}\dfrac{\gamma_R}{\gamma_T}\sqrt{\dfrac{L_T}{L_R}} & \dfrac{\omega}{\omega_R} - j\gamma_R - \dfrac{\omega_R}{\omega}\dfrac{1}{(1-k_{TR}^2)} \end{bmatrix} \begin{bmatrix} I_N \\ I_L \end{bmatrix} = 0 \quad (8)$$

The structural parameters of the transmitting end are the same: $L_1 = L_3 = L$, $C_1 = C_3 = C$ , equivalent inductance at the transmitting end: $L_T = L_1 L_3 / (L_1 + L_3) = L/2$ , equivalent capacitance at the transmitting end: $C_T = C_1 + C_3 = 2C$ , equivalent inductance and capacitance at the receiving end: $L_R = L_2$, $C_R = C_2$ .Transmitting end loss: $\gamma_{\mathrm{T}} = \sqrt{L_T / C_T} / R_{\mathrm{N}}$ ,

<

receiving end loss: $\gamma_{\mathrm{R}} = \sqrt{L_R / C_R} / R_{\mathrm{L}}$ , the coupling coefficient between the transmitting coil and the receiving coil is $k_i = M_{i2} / (L_i L_{\mathrm{R}})^{1/2} (i = 1,2,3)$ ,Transmitting end equivalent coupling coefficient: $k_{PS} = k_1 (L_T / L_1)^{1/2} + k_3 (L_T / L_3)^{1/2}$ and $k_{TR} = (k_1^2 + k_3^2)^{1/2}$ , When $k_1$ and $k_2$ are equal or similar and small, then there is $k_{PS}^2 \approx k_{TR}^2$ , Ordering $\omega_1 = 1/(L_1 C_1)^{1/2}$ , $\omega_2 = 1/(L_2 C_2)^{1/2}$ , $\omega_3 = 1/(L_3 C_3)^{1/2}$ as the natural resonant angular frequencies formed by each resonant cavity, the overall resonant frequency of the transmitting end is $\omega_T = 1/(L_T C_T)^{1/2}$ , the overall resonant frequency of the receiving end is $\omega_R = 1/(L_R C_R)^{1/2}$ .

In order to ensure the stable and efficient transmission of energy at the receiving and transmitting ends of the system, it is necessary to make the resonant angular frequencies equal in order to achieve the effect of magnetic coupling resonance. In addition, in order for this system to achieve PT symmetry, the following conditions must be met:

$$\begin{cases} \omega_1 = \omega_2 = \omega_T = \omega_{\mathrm{R}} = \omega_0 \\ \dfrac{1}{R_{\mathrm{N}}} \sqrt{\dfrac{L_{\mathrm{T}}}{C_{\mathrm{T}}}} = \dfrac{1}{R_{\mathrm{L}}} \sqrt{\dfrac{L_{\mathrm{R}}}{C_{\mathrm{R}}}} \end{cases} \tag{9}$$

From the above equation 9, it can be concluded that: $L_R = 2L_T = L$ , $C_R = C_T / 2 = C$ , $2R_N = R_L$ ,satisfying the PT symmetry condition.

Find the expression for $\omega$ :

$$\omega_{1,2} = \sqrt{\frac{2-(1-k_{\mathrm{TR}}^2)\gamma_{\mathrm{R}}^2 \pm \sqrt{(2-(1-k_{\mathrm{TR}}^2)\gamma_{\mathrm{R}}^2)^2 - 4(1-k_{\mathrm{TR}}^2)}}{2LC(1-k_{\mathrm{TR}}^2)}} \tag{10}$$

Using quality factor here $Q = R_L (C_R / L_R)^{1/2}$ and the equivalent coupling coefficient $k = (k_1^2 + k_3^2)^{1/2}$ , and substituting them into equation 10 above:

$$\omega_{1,2} = \sqrt{\frac{2Q^2 - 1 + k^2 \pm \sqrt{(2Q^2 - 1 + k^2)^2 - 4Q^4(1-k^2)}}{2LCQ^2(1-k^2)}} \tag{11}$$

The expression for the critical coupling coefficient $k_{C2}$ is:

$$k_{C2} = \sqrt{1 - 2Q^2 - 2Q^4 + 2Q^3\sqrt{Q^2 + 2}} \tag{12}$$

This critical coupling coefficient expression is consistent with that in a one-to-one system (equation 5), indicating that the two-to-one system structure does not affect the critical coupling coefficient. At this point, substituting the expression for $\omega$ into the equation yields: $|I_L|/|I_N| = \sqrt{L_T / L_R}$ .

As $L_R = 2L_T = L$, $C_{\mathrm{R}} = C_{\mathrm{T}} / 2 = C$ , $2R_N = R_L$ :

$$\frac{U_L}{U_i} = \frac{|I_L| R_L}{|I_N| R_N} = \sqrt{2} \tag{13}$$

$U_L$ and $U_i$ are the voltages at the receiving and transmitting ends, respectively. The theoretical transmission efficiency is :

$$\eta = \frac{P_L}{P_{\mathrm{i}}} = \frac{U_L^2 R_N}{U_i^2 R_L} = 100\% \tag{14}$$

From this, it can be seen that the system can increase the voltage at the receiving end while maintaining the same transmission efficiency.

Another case is the relatively simple two-to-one system B. In this system, the receiving coil B has a larger size-twice that of the transmitting coils—enabling it to couple with double-transmitting coils simultaneously. This is equivalent to having double-transmitting systems coupled and operating with the same receiving system at the same time. As shown in Figure 3. Modeling and analyzing the system, ignoring the influence of

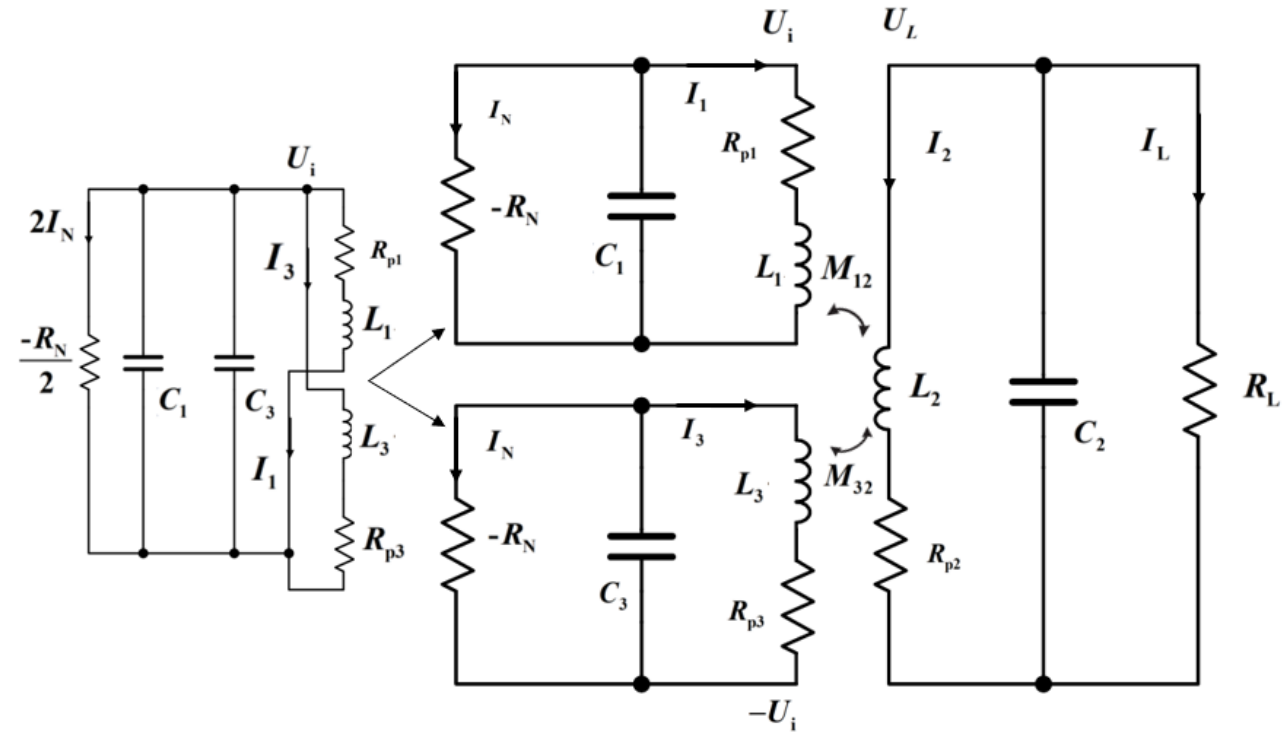

**Fig. 3.** Equivalent circuit diagram of two-to-one system B

coil internal resistance, using Kirchhoff's voltage and current laws, we can obtain equation 15:

$$\begin{gathered} U_{\mathrm{i}} = j\omega L_1 I_1 + j\omega M_{12} I_2 \\ U_{\mathrm{i}} = j\omega L_3 I_3 + j\omega M_{32} I_2 \\ U_L = j\omega M_{12} I_1 + j\omega M_{32} I_3 + j\omega L_2 I_2 \\ I_1 + I_3 + j\omega(C_1 + C_3) U_{\mathrm{i}} - \frac{2U_{\mathrm{i}}}{R_N} = 0 \\ I_2 + j\omega C_2 U_L + \frac{U_L}{R_L} = 0 \end{gathered} \tag{15}$$

Due to the symmetry of the transmitting circuit, it can be concluded that: $L_1 = L_3 = L$ , $I_1 = I_3 = I$ , $C_1 = C_3 = C$ . And at this point, assuming that the coupling coefficients of the double-transmitting coils and the receiving coil are the same, then there is: $M_{12} = M_{32} = M$ .

Converted into matrix form:

$$\begin{bmatrix} 1 - \omega^2 LC - j\dfrac{\omega L}{R_N} & -\omega^2 MC_2 + j\dfrac{\omega M}{R_L} \\ -2\omega^2 MC - 2j\dfrac{\omega M}{R_N} & 1 - \omega^2 L_2 C_2 + j\dfrac{\omega L_2}{R_L} \end{bmatrix} \begin{bmatrix} U_i \\ U_L \end{bmatrix} = 0 \tag{16}$$

When the following conditions are met, use the matrix to find the solution: $L_2 = L, \quad C_2 = C, \quad 2R_N = R_L$ .

The expression for $\omega$ is:

$$\omega_{1,2} = \sqrt{\frac{LC - \dfrac{L^2 - 2M^2}{R_L^2} \pm \sqrt{\left(LC - \dfrac{L^2 - 2M^2}{R_L^2}\right)^2 - C^2(L^2 - 2M^2)}}{C^2(L^2 - 2M^2)}} \tag{17}$$

Replacing with $k = M / L$ and $Q = R_{\mathrm{L}} (C / L)^{1/2}$ yields:

$$\omega_{1,2} = \sqrt{\frac{Q^2 - 1 + 2k^2 \pm \sqrt{\left(Q^2 - 1 + 2k^2\right)^2 - Q^4(1-2k^2)}}{Q^2 LC(1-2k^2)}} \tag{18}$$

The critical coupling coefficient $k_{C3}$ is:

$$k_{C3} = \sqrt{-\frac{1}{4}Q^4 - \frac{1}{2}Q^2 + \frac{1}{2} + \frac{1}{4}Q^3\sqrt{Q^2+4}} \tag{19}$$

In this case, since both coils at the transmitting end are coupled to the receiving end with the same coupling coefficient, it can be considered that two negative resistance structures are simultaneously coupled to the receiving end. At this time, the voltage ratio between the two ends is:

$$\frac{U_L}{U_\mathrm{i}} = \frac{|I_L|\,R_L}{|I_N|\,R_N} = \sqrt{\frac{L_T}{L_R}\frac{R_L}{R_N}} = 2 \tag{20}$$

$$\eta = \frac{P_L}{P_\mathrm{i}} = \frac{{U_\mathrm{L}}^2 R_N}{2{U_\mathrm{i}}^2 R_L} = 100\% \tag{21}$$

## III. ANALYSISANDCOMPARISONOFTRANSMISSION CHARACTERISTICS

The size parameters of the transmitting coils in both systems are the same. Since the critical coupling coefficient and frequency expression of the two-to-one receiving coil A system are the same as those of the one-to-one system, the one-to-one system will not be analyzed here. Only the two-to-one receiving coil A and the two-to-one receiving coil B will be examined. Based on equations 12 and 19, along with $Q = R_\mathrm{L}(C/L)^{1/2}$, we obtained Figures 4 and 5.

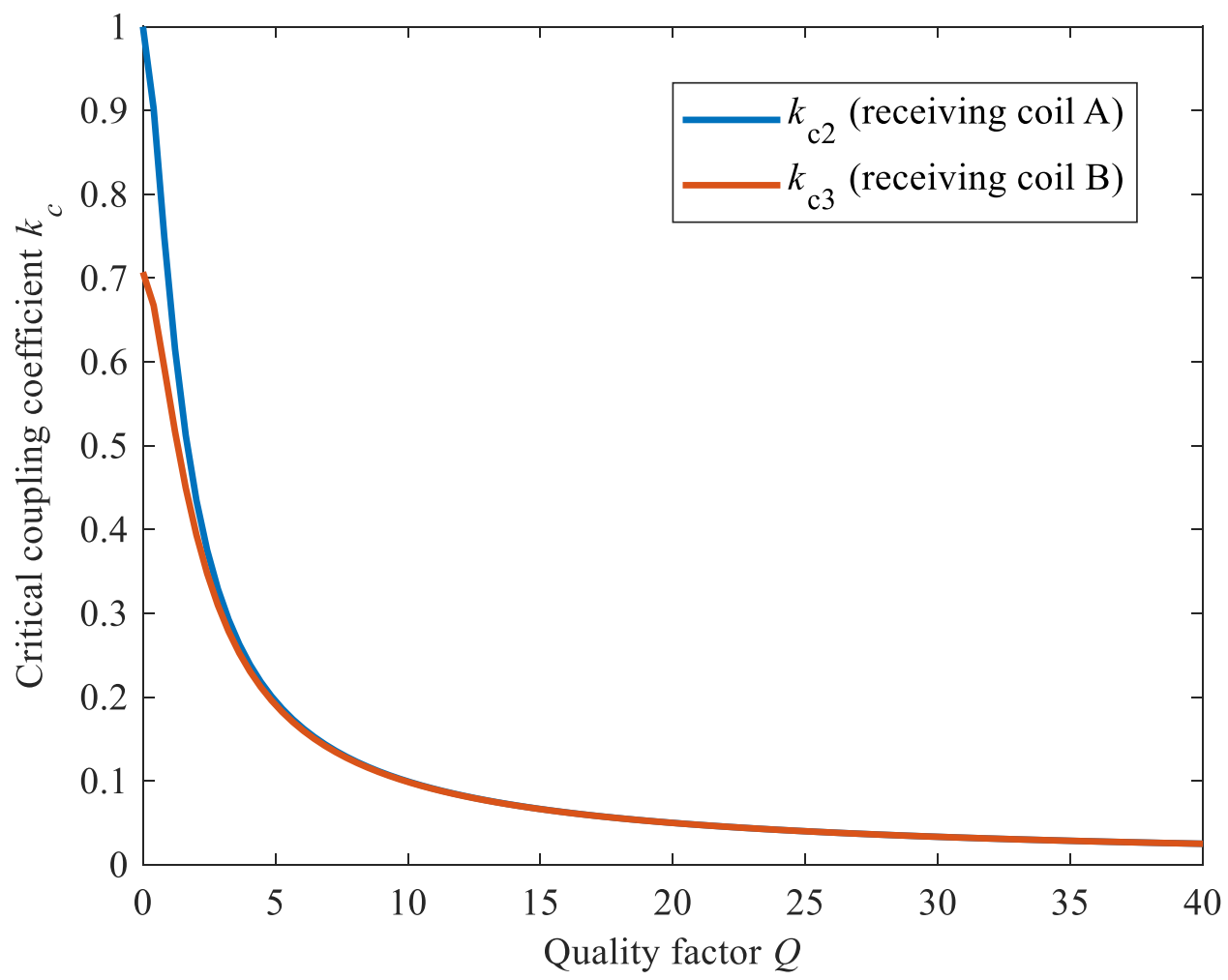

**Fig. 4.** The trend of critical coupling coefficient changing with quality factor.

From Figure 4, it can be seen that system B achieves a reduction in the critical coupling coefficient, which translates to an increase in transmission distance. However, when the quality factor is high, the extent of this reduction is relatively. Figure 5, under conditions where the load resistance remains minor, and consequently, its impact on extending transmission distance is not significant in such cases, the transmission distances of the two systems are approximately equal. Due to the linear relationship between quality factor $Q$ and load resistance $R_\mathrm{L}$, the reduction trend of Figure 5 is similar to Figure 4. In summary, when the $Q$ or $R_\mathrm{L}$ is large enough, the critical coupling coefficient is minor for a long-distance transfer.

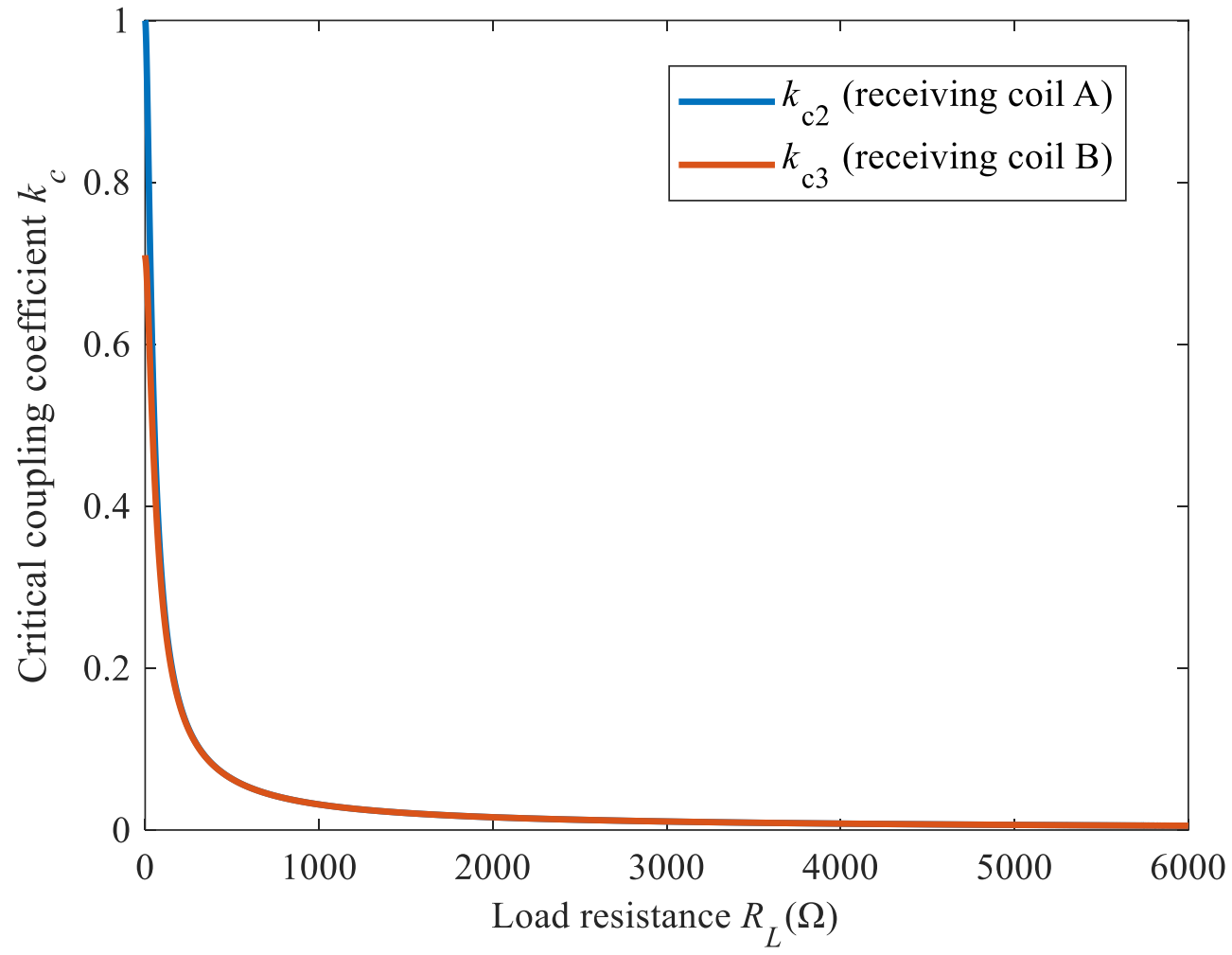

**Fig. 5.** The trend of critical coupling coefficient changing with load resistance.

## IV. EXPERIMENTAL VERIFICATION

### *A. Implementation of Negative Resistance*

LM6171 voltage amplifier is selected as the negative resistor to form a negative resistor circuit, as shown in Figure 6. Table I shows component parameters of the systems.

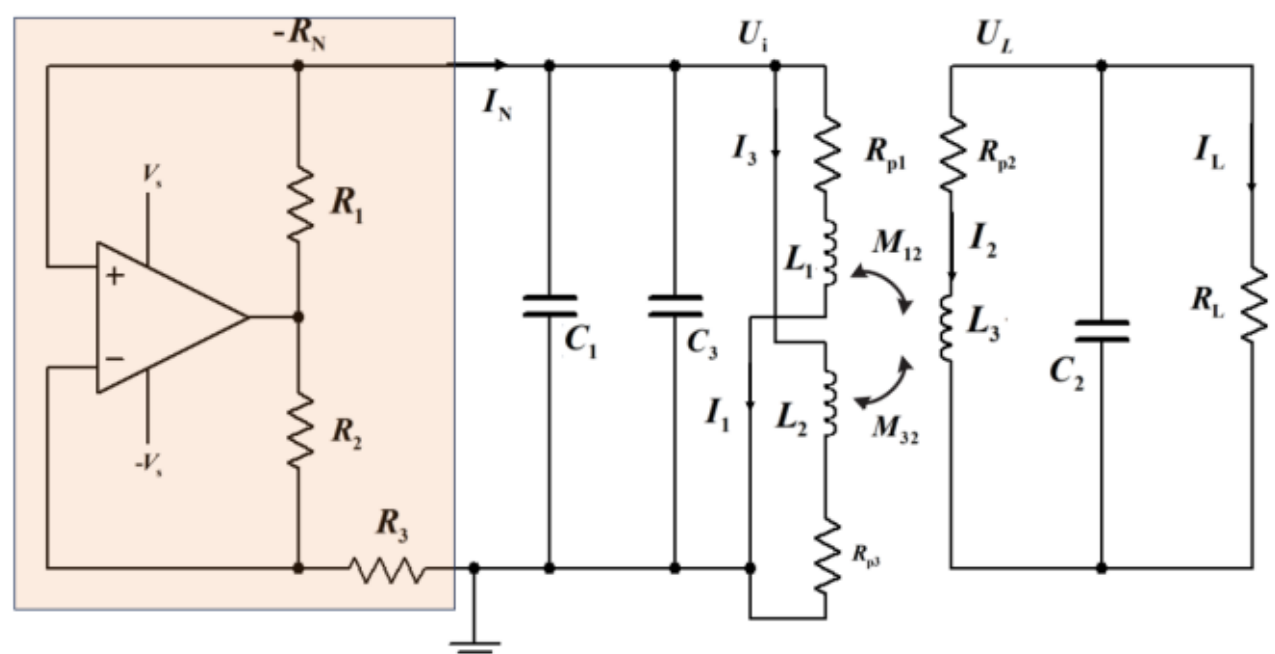

**Fig. 6.** A circuit structure composed of negative resistors

When the OA operates in the linear working region, the relationship between the input and output currents can be derived from the virtual break and virtual short characteristics, and the equivalent resistance, that is, the negative resistance, can be derived as $-R_N = -R_1R_3/R_2$. When the OA circuit provides energy to the outside, the OA circuit is equivalent to the negative resistance providing energy to the positive resistance circuit.

Previous studies have typically OA in linear operating regions, his is only applicable to one-to-one systems, thus presenting a limitation[12] [14] [15]. Therefore, a new method for composing negative resistance circuits based on OA is proposed. $R_2$ is configured to be ten times the resistance of $R_3$ and the amplification factor A of the OA is maintained at a high gain (A=10), allowing the OA to operate in a nonlinear state. When the OA oscillates, the voltage at the reverse input terminal is small, while the output voltage at the same input terminal is high. The OA no longer satisfies the virtual short virtual break characteristic and operates in the forward saturation region, outputting energy outward. This circuit has stronger stability.

Table I: Component Parameters of the System

| Symbol | Values |
|---|---|
| $L_1$, $L_2$ ,$L_3$ | $60uH$ |
| $C_1$, $C_2$ ,$C_3$ | $60nF$ |
| $R_1$,$R_3$ | $51\Omega$ |
| $R_2$ | $510\Omega$ |
| $R_L$ | $5600\Omega$ |
| $R_{p1}$, $R_{p2}$, $R_{p3}$ | $0.5\Omega$ |

*B. Experimental Setup*

The experimental voltage source is selected as a DC stabilized voltage source, with a supply voltage of ± 15V. The transmitting and receiving coils are both wound with Litz wire (0.01mm×200), with an outer diameter of 140 mm. The receiving coil A is the same size as the transmitting coil, with a size of 140 mm, and the receiving coil B is 290 mm. The two transmitting coils are placed flat together without overlapping, and the measured voltage varies with vertical distance and horizontal offset position. The experimental testing site is shown in Figure 7.

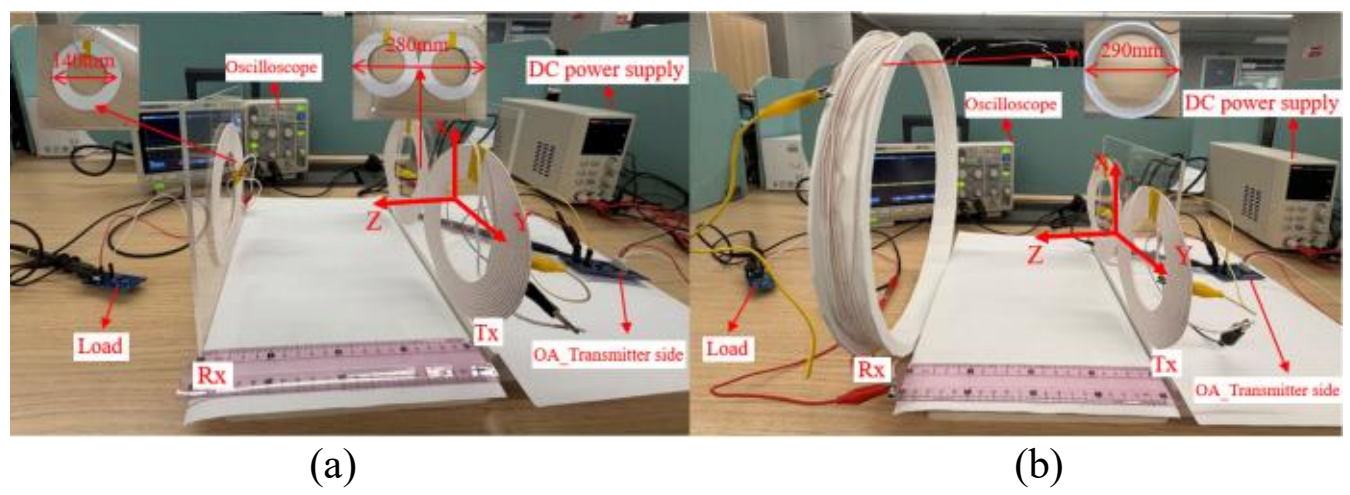

**Fig. 7.** Experimental prototype of two-to-one system. (a)receiving coil A. (b)receiving coil B.

*C. Experimental Results*

1. Transmission characteristics of receiving coil A.

The size of receiving coil A is 140 mm, used to verify the performance of the PT symmetry system. From Figure 8, it can be seen that in the vertical Z direction, in the strongly coupled symmetric region, the voltage values at the receiving and transmitting ends are not significantly different from the theoretical values. In the strongly coupled region of PT symmetry, at around 12 cm, the coupling coefficient reaches the critical coupling coefficient value 0.0057 (equation 12), and the PT symmetry condition is broken, causing the voltage to begin to drop. From Figure 8, it can be seen that the theoretical voltage ratio has been close to 1.41 in the strongly coupled symmetry region, while the actual test value is above 1.37, which is the result of various parasitic parameters affecting it.

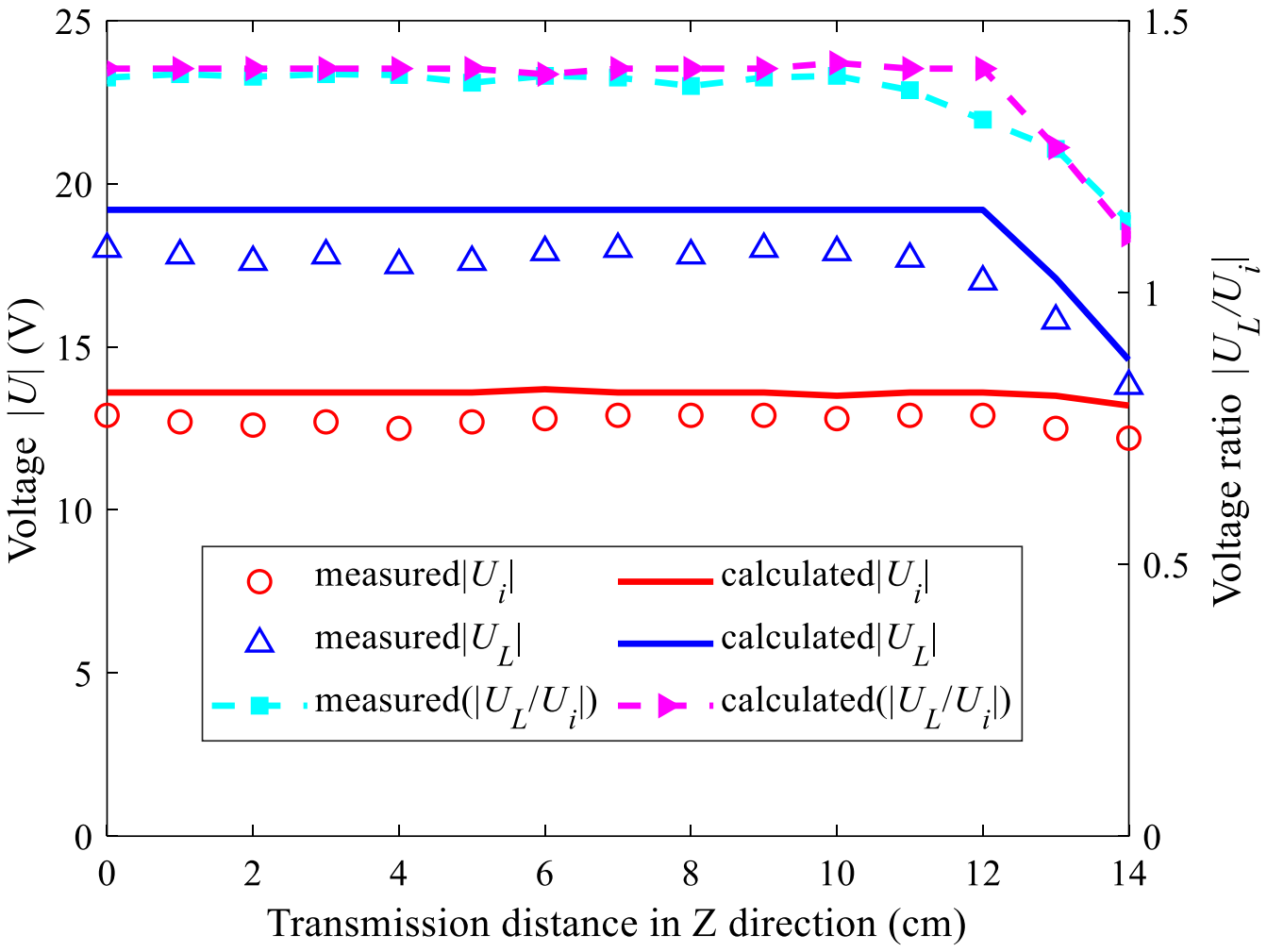

**Fig. 8.** Z direction voltage result (receiving coil A).

Below, we horizontally move the receiving coil in the Y direction at Z = 4 cm and observe the change in the receiving voltage. Moving the coil in this direction essentially results in a change in the equivalent coupling coefficient after the system enters the strong coupling symmetry region, which is equivalent to a vertical movement in the Z direction. It can be seen that when Y is greater than 15 cm, the system enters the weak coupling, that is, the asymmetric region as shown in Figure 9.

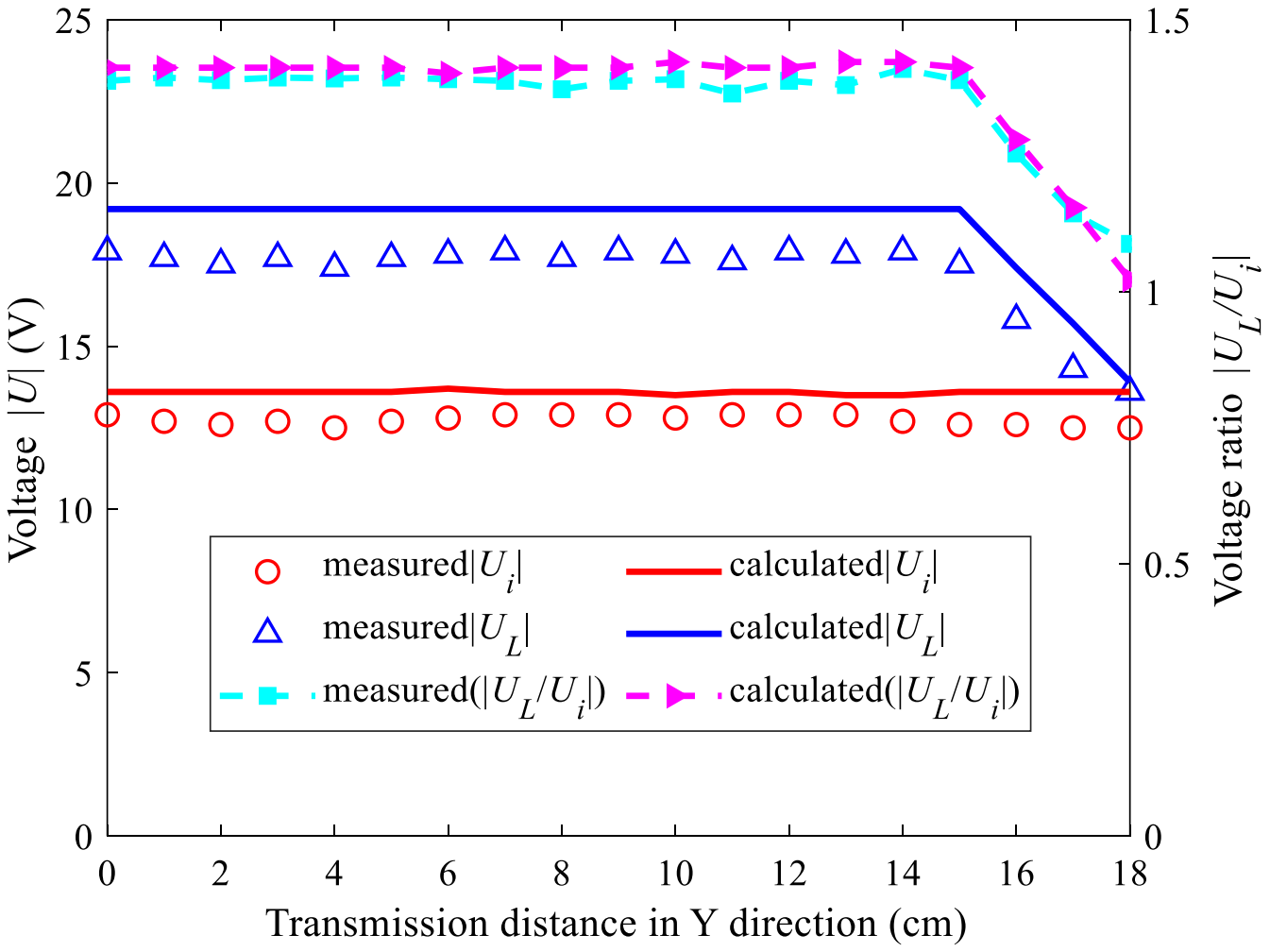

**Fig. 9.** Y direction voltage result (receiving coil A).

Similarly, at Z = 4 cm, we horizontally move the receiving coil in the X direction and observe the change in the receiving voltage. Moving the coil in this direction essentially results in a change in the equivalent coupling coefficient after the system enters the strong coupling symmetry region, which is equivalent to a vertical movement in the Z direction. When moving in the X direction, the equivalent coupling coefficient decreases faster. As shown in Figure 10, when X is greater than 7 cm, the system enters the weak coupling asymmetry region.

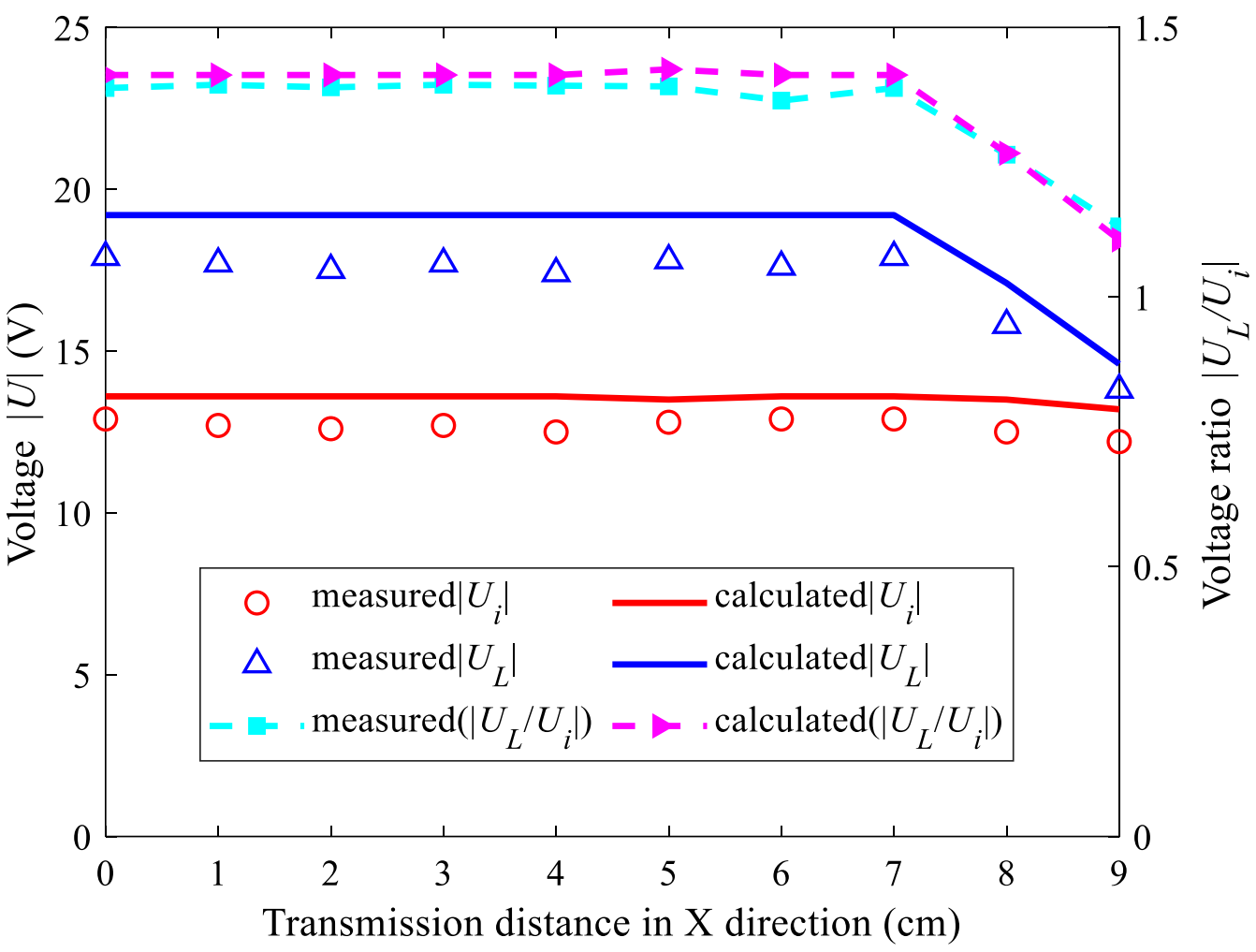

**Fig. 10.** X direction voltage result (receiving coil A).

When moving vertically in the Z direction, stable voltage output can be maintained within 12 cm. In the Y direction, stable voltage output can be maintained with a 15 cm offset distance. In the X direction, stable voltage output can be maintained with a 7 cm offset distance. The changes in the three distances are equivalent coupling coefficient changes in the strong coupling region. In the PT symmetry region, the maximum measured voltage at the receiving end is 18.0 V, the minimum measured value is 17.5 V, and the voltage difference is 0.5 V. The highest and lowest measured voltages at the transmitting end are 12.9 V and 12.5 V, and the voltage difference is 0.4 V. The fluctuation rate of the receiving end voltage does not exceed 2.8%. The theoretical voltage ratio is close to 1.41 in the strong coupling symmetry region, and the actual test value is above 1.37. The actual transmission efficiency remains above 94.4%. Therefore, this system can achieve misalignment tolerance of horizontal offset when moving vertically in the coil. Ability, Figure 11 shows the curve of the coupling coefficient as a function of operating frequency.

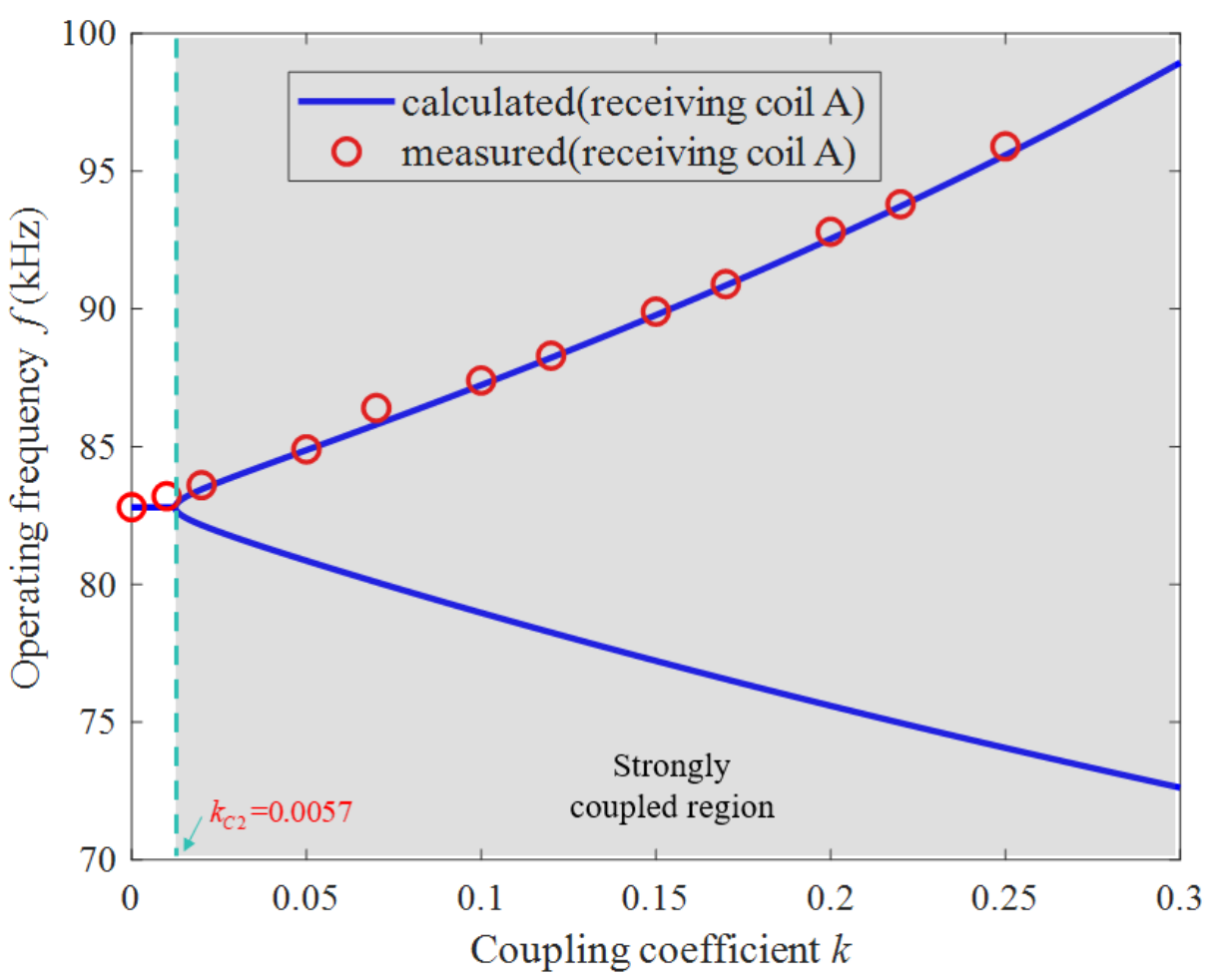

**Fig. 11.** Operating frequency in system A as a function of coupling coefficient.

2. Transmission characteristics of receiving coil B

The receiving coil B has a diameter of 290 mm and is used to verify the performance of the PT symmetry system. As shown in Figure 12, along the vertical Z direction within the strongly coupled symmetric region, the measured voltage values at the receiving and transmitting ends show good agreement with the theoretical predictions. The PT symmetric strongly coupled region occurs at approximately 14 cm, where the coupling coefficient reaches the critical value of 0.0056 (equation 19). Beyond this point, the PT symmetry condition breaks down, leading to a rapid decline in voltage.

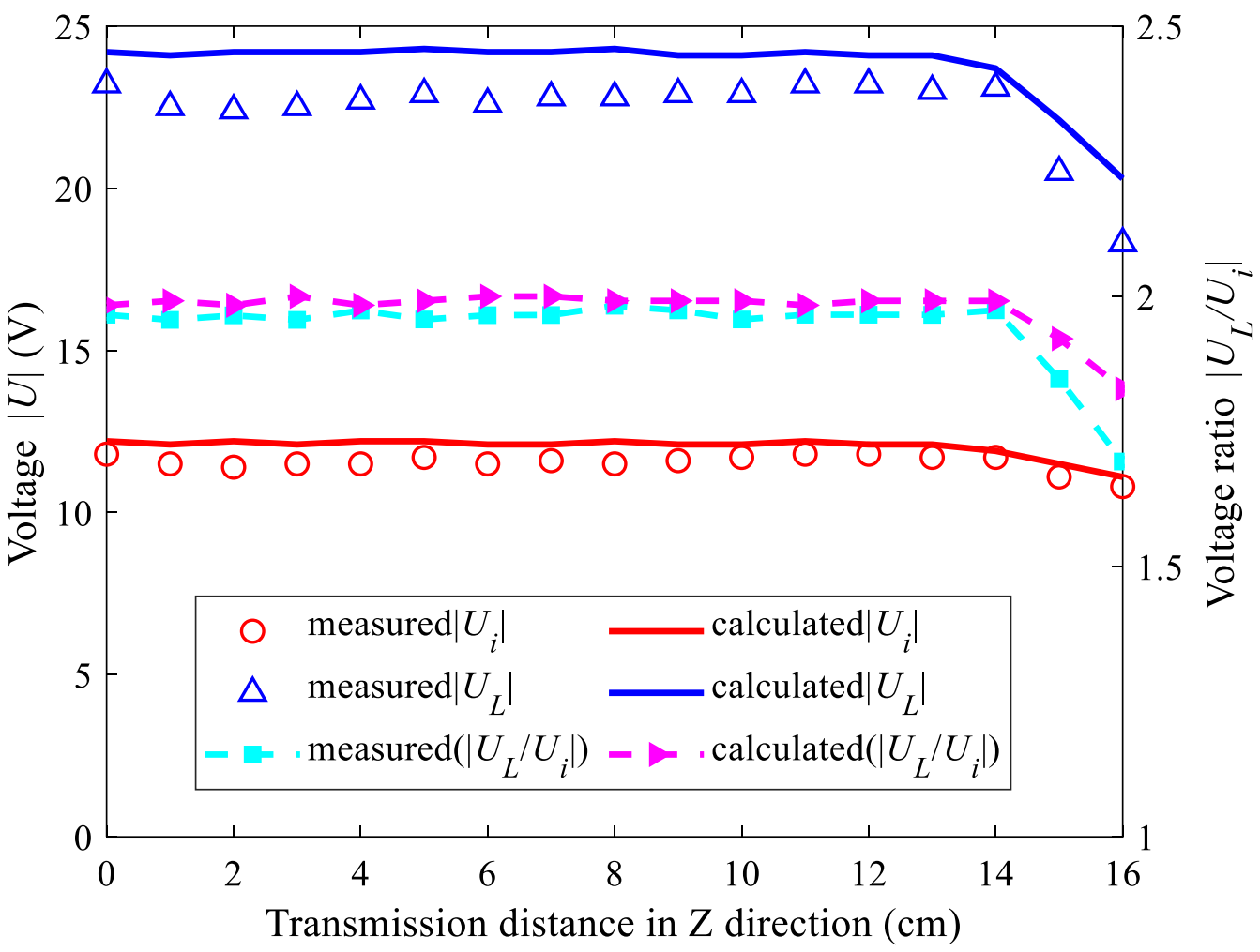

**Fig. 12.** Z direction voltage result (receiving coil B).

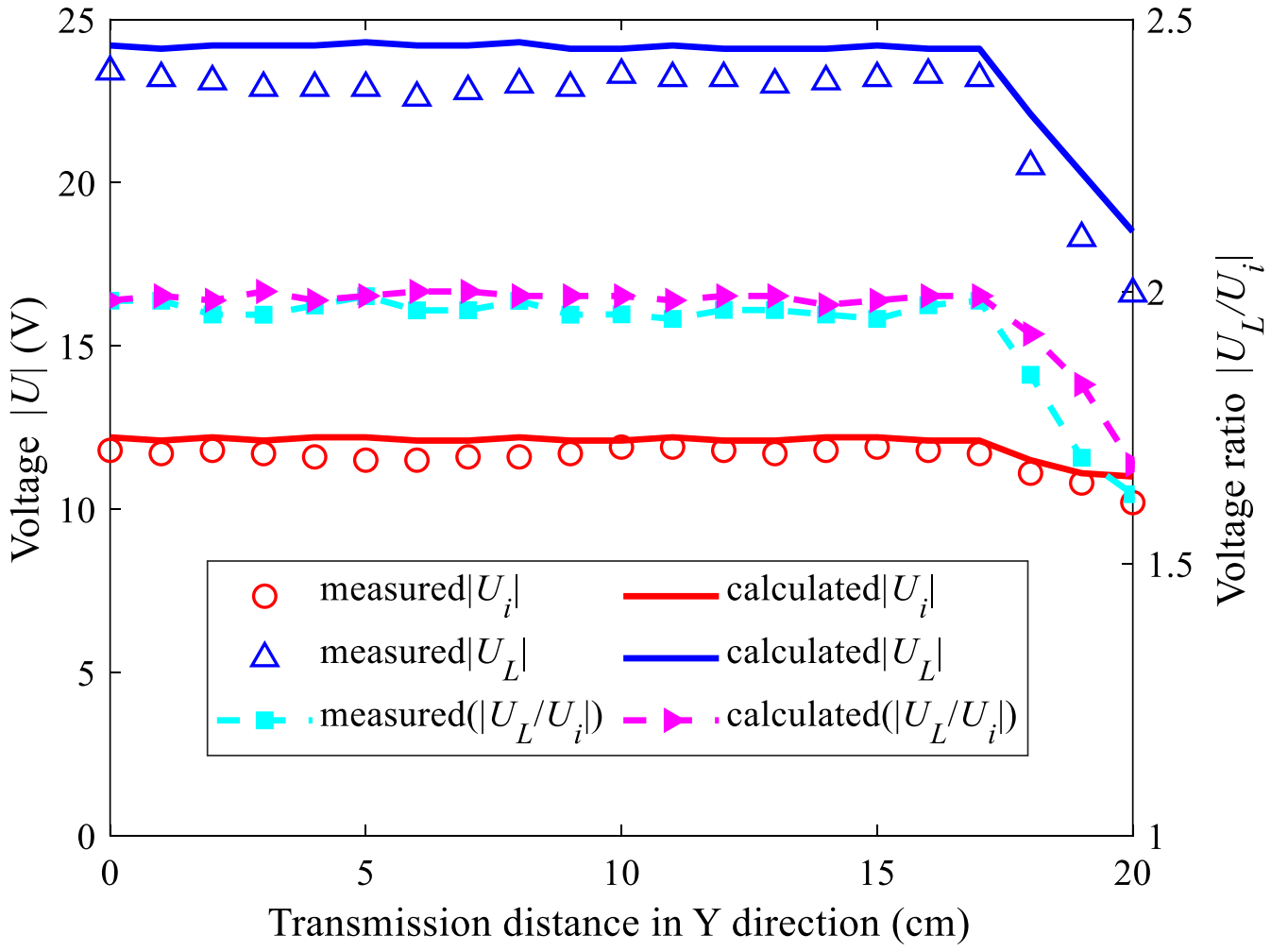

**Fig. 13.** Y direction voltage result (receiving coil B).

Next, we horizontally move the receiving coil in the Y direction at Z=4 cm and observe the change in receiving voltage. In the strong coupling symmetry region, moving the coil in this direction is essentially a change in equivalent coupling coefficient, equivalent to vertical movement in the Z direction. As shown in Figure 13, when Y is greater than 17 cm, the system enters the weak coupling PT asymmetry region.

<

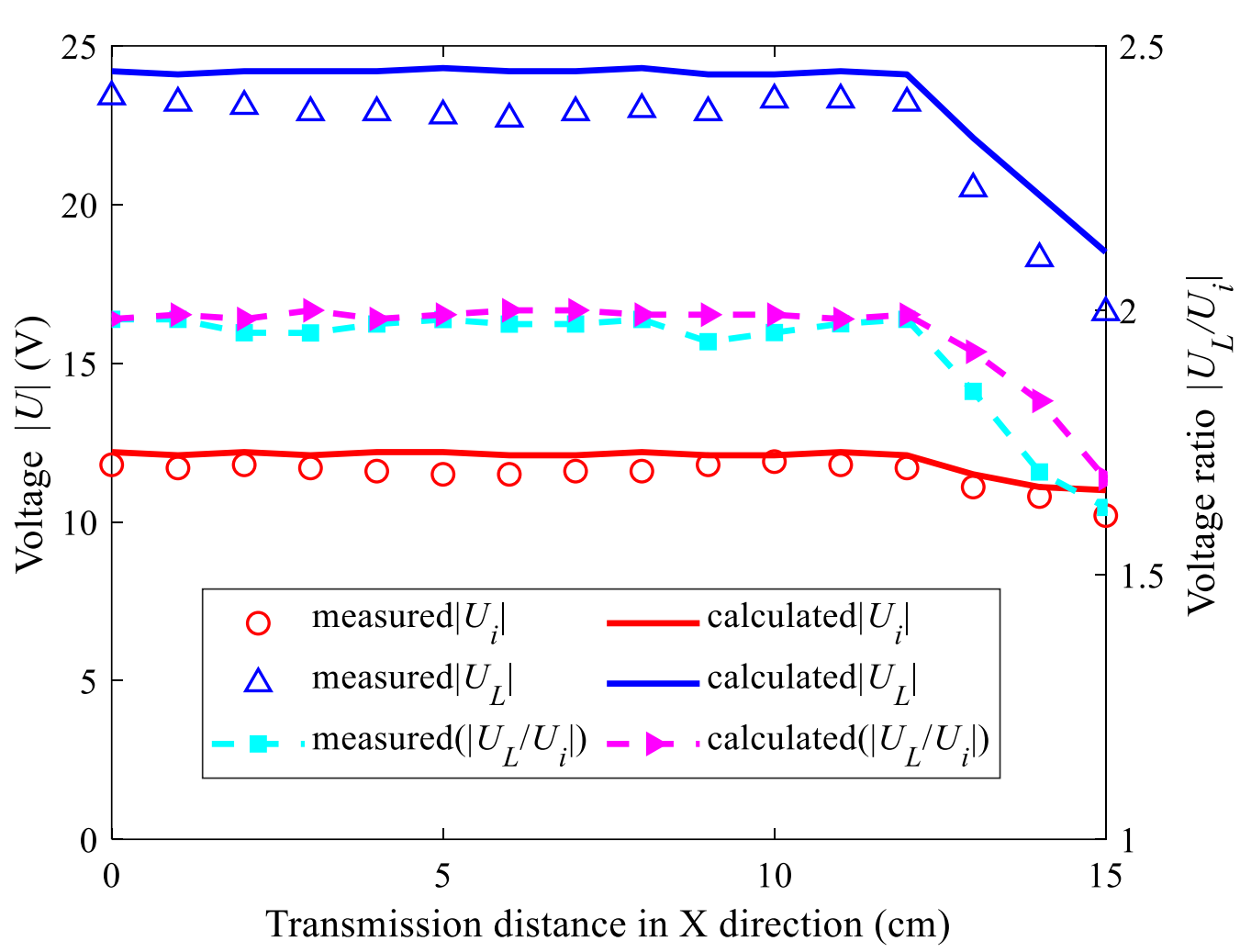

**Fig. 14.** X direction voltage result (receiving coil B).

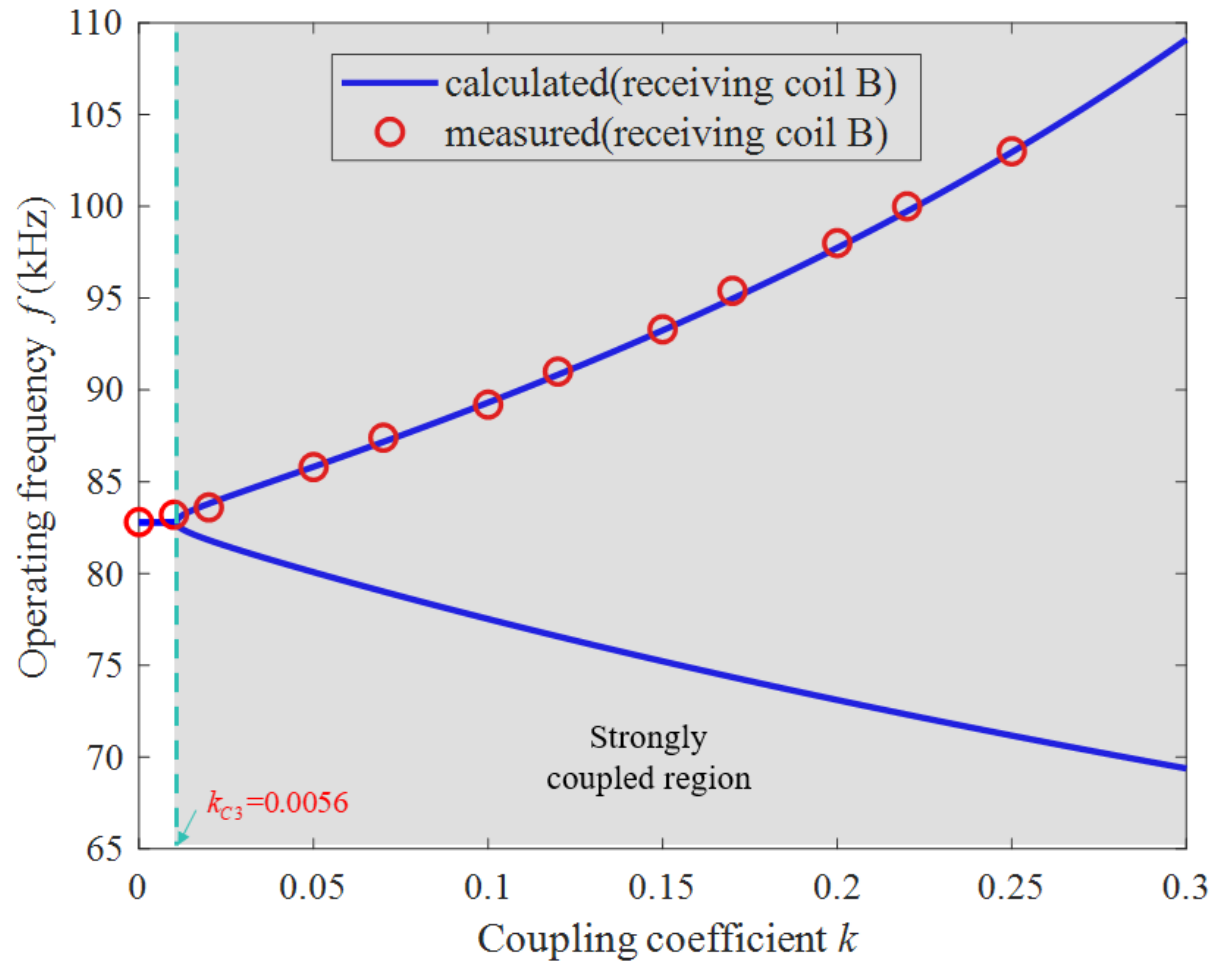

**Fig. 15.** Operating frequency in system B as a function of coupling coefficient.

Similarly, as shown in Figure 14, at Z=4 cm, we horizontally move the receiving coil in the X direction and observe the change in the receiving voltage. Moving the coil in this direction essentially results in a change in the equivalent coupling coefficient after the system enters the strong coupling symmetry region, which is equivalent to a vertical movement in the Z direction. When moving in the X direction, the equivalent coupling coefficient decreases faster. It can be seen that when X is greater than 13 cm, the system enters the weak coupling, that is, the asymmetric region.

When moving vertically in the Z direction, the system can maintain a stable voltage output within a range of 14 cm. In the Y direction, stable output is preserved with an offset distance of up to 17 cm, while in the X direction, the offset tolerance is 13 cm. These variations in displacement correspond to changes in the equivalent coupling coefficient within the strong coupling region. Within the PT symmetry region, the measured maximum and minimum voltages at the receiving end are 23.2 V and 22.4 V, respectively, resulting in a voltage difference of 0.8 V. At the transmitting end, the maximum and minimum voltages are 11.8 V and 11.5 V, respectively, yielding a voltage difference of 0.3 V.

The receiving voltage fluctuation rate remains below 3.4%. The theoretical ratio of received to transmitted voltage remains close to 2 in the strongly coupled symmetric region, with actual test values consistently above 1.95. The actual transmission efficiency remains above 95.1%. Hence, the system exhibits robustness against horizontal misalignment while the coil is moving along the vertical direction. Figure 15 presents the curve of operating frequency versus coupling coefficient.

Finally, we present voltage waveform diagrams of two receiving structures in three directions, as shown in Figure 16.To further illustrate the advantages of voltage boost in the two-to-one systems A and B, Figure 17 shows a comparison of the received voltage values of the two-to-one systems A and B, as well as the one-to-one system . $U_A$ represents the voltage at the receiving end of system A, $U_B$ represents the voltage at the receiving end of system B, and $U_C$ represents the voltage at receiving end of one-to-one system.

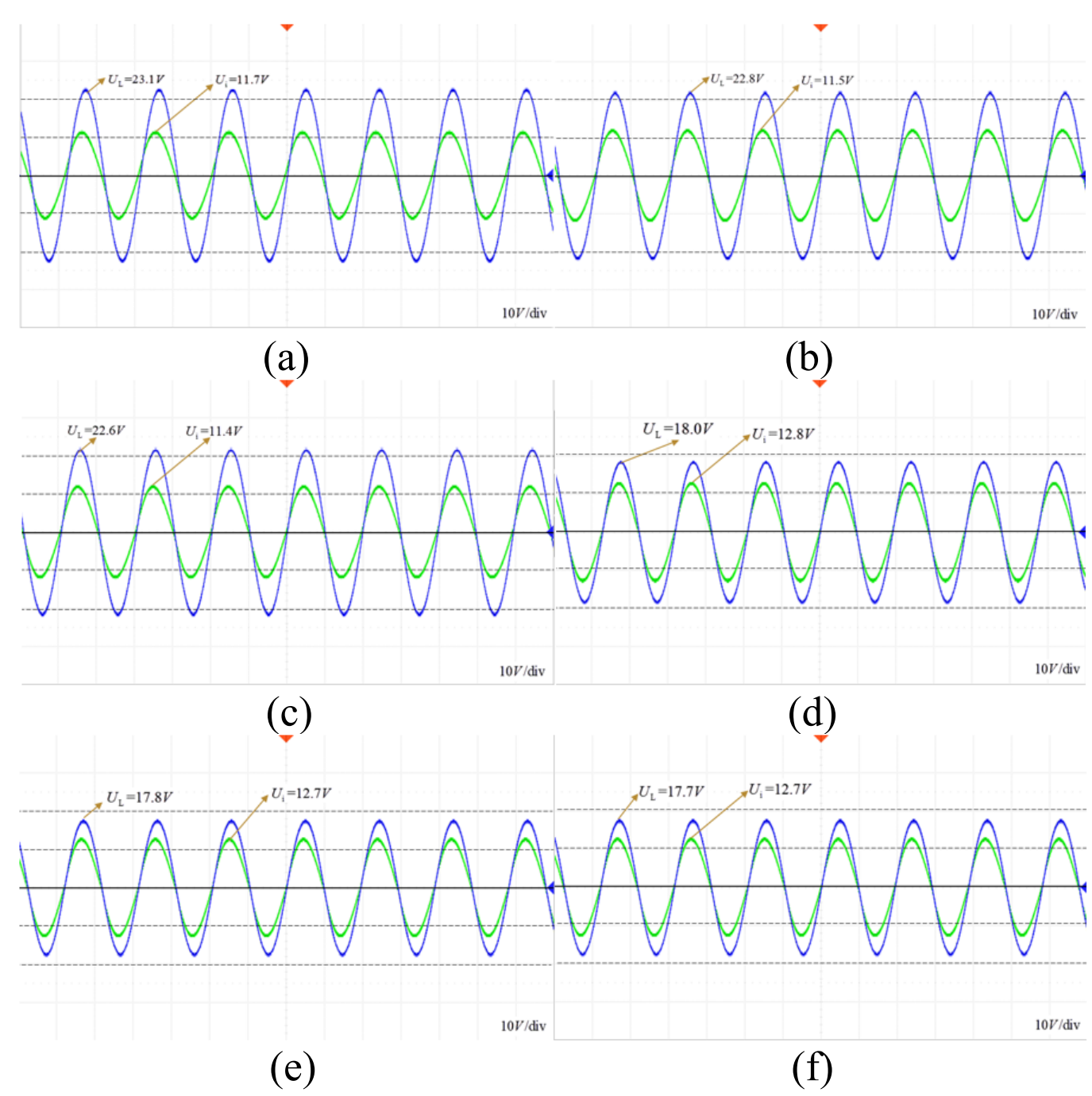

**Fig. 16.** Actual voltage test waveform. (a) Z direction (system B). (b) Y direction (system B). (c) X direction (system B). (d) Z direction (system A). (e) Y direction (system A). (f) X direction (system A).

The transmitting and receiving coils used in one-to-one system are consistent with those used in system A. From Figure 17, it can be seen that the voltage of the two-to-one system A has increased to about 1.36 times compared to the one-to-one system , the voltage of the two-to-one system B has increased to about 1.77 times compared to the one-to-one system, and the voltage of the two-to-one system B has increased to about 1.28 times compared to the two-to-one system A. This is primarily because in the two-to-one system, the PP configuration of the double-transmitting coil reduces the voltage at the transmitting end, resulting in a slightly lower transmitting voltage compared to the one-to-one system.

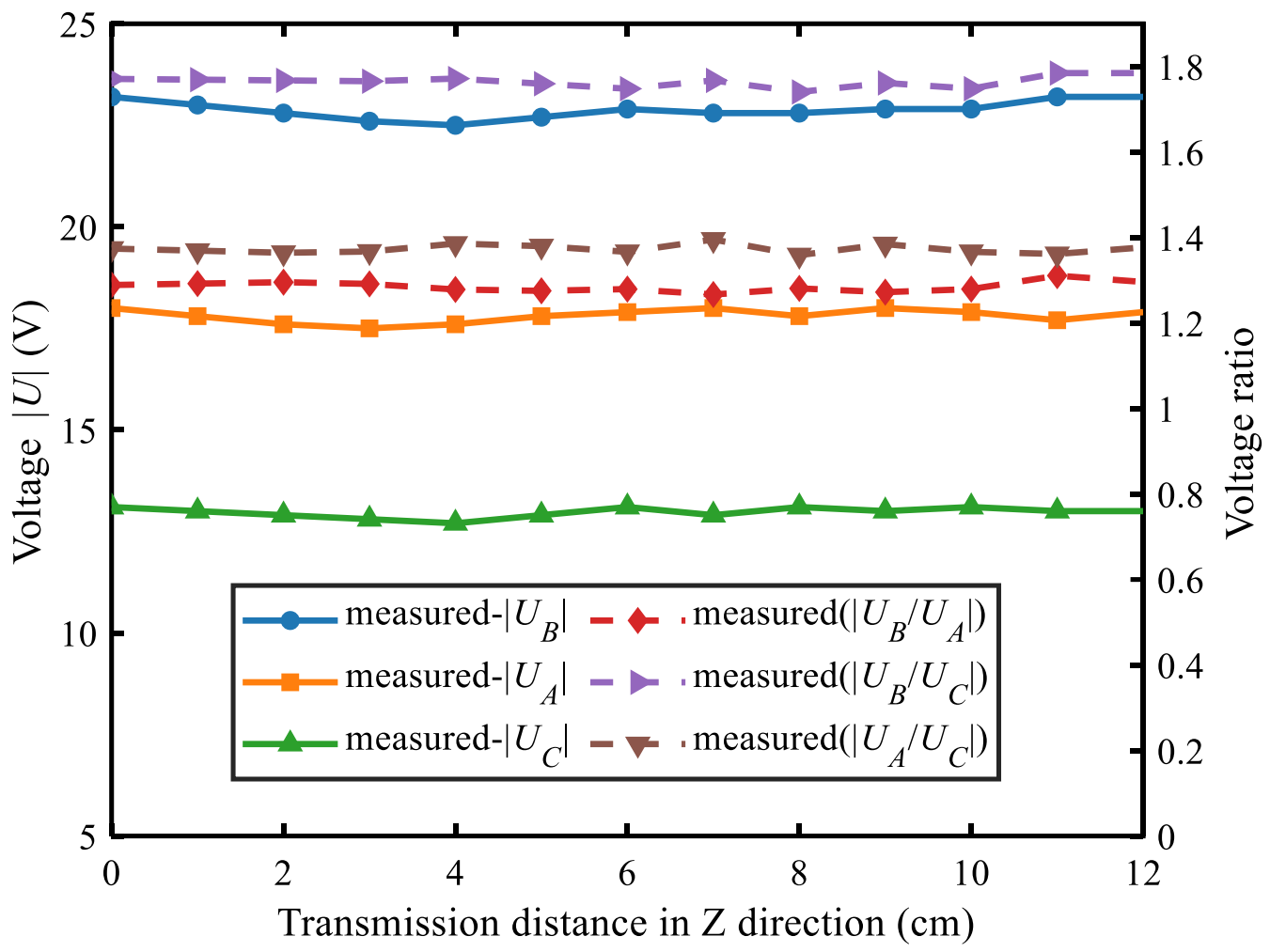

**Fig. 17.** Comparison results of three systems in Z direction.

Next, as shown in Figure 18, we compared the output power of the three systems with the variation trend of load resistance. The load resistance changed from 50 ohms to 8000 ohms. Since three systems adopt a circuit structure in which capacitor and inductor are connected in parallel, this structure performs better under high load resistance. At low load resistance, the transmitting terminal voltage cannot reach the highest output voltage, and the transmission distance is low. However, the output power is high, and the highest output power can reach 510 mW. At high load resistance (above 2000 ohms), the output voltage reaches the highest voltage, and the output voltage remains unchanged. The larger the load resistance, the lower the output power, and the farther the transmission distance. Compared to one-to-one system, two-to-one systems can increase the output power by increasing the load receiving terminal voltage.

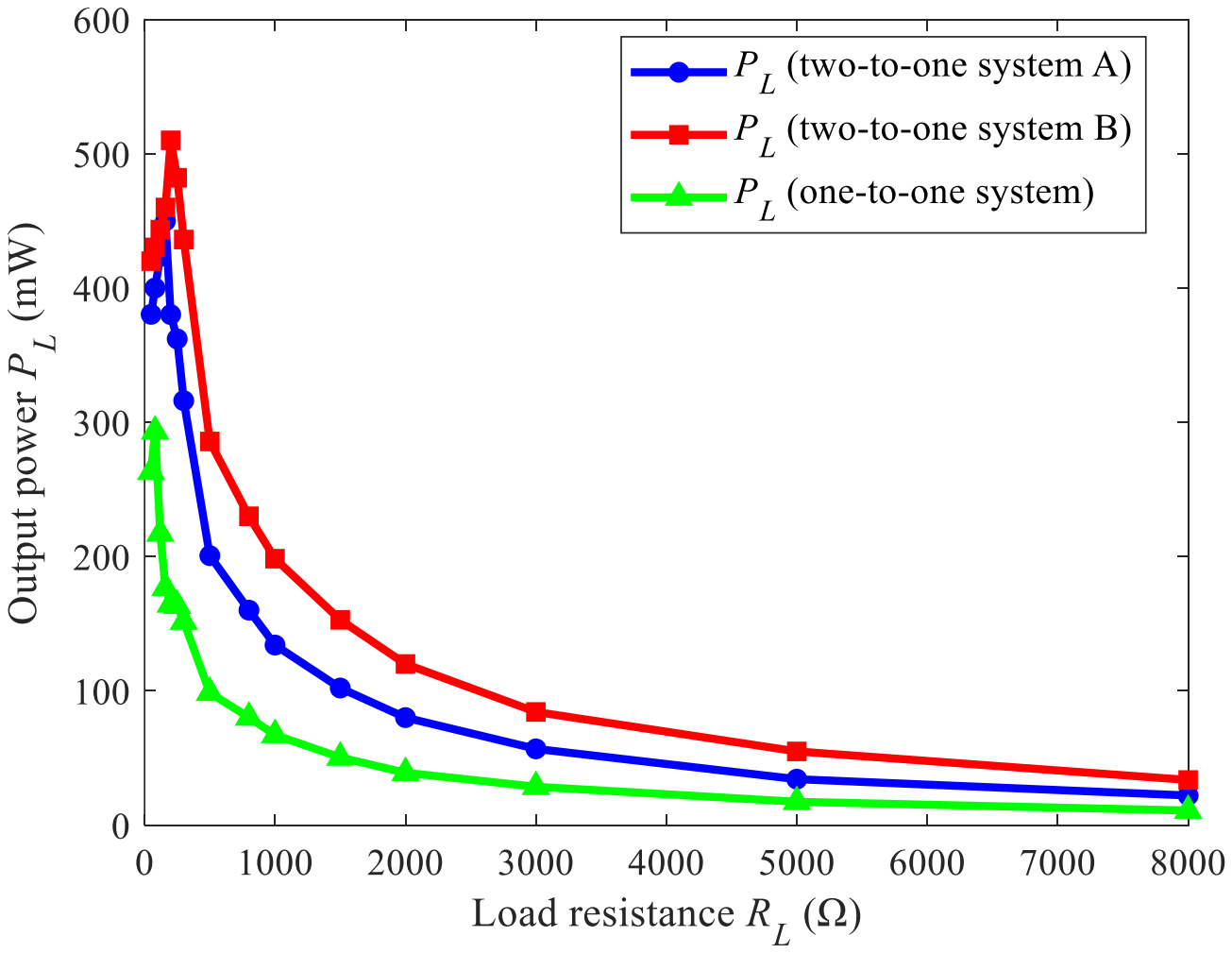

**Fig. 18.** Three types of system output power variation curves with load resistance.

This chapter fully verifies the performance of PT symmetry WPT systems through detailed experiments. Experimental results have shown that under strong coupling symmetry working conditions, the system can still maintain high power stable output in response to changes in coupling coefficients introduced by three-dimensional spatial displacement (XYZ), greatly improving the robustness during spatial position changes and increasing power. Comparing to the one-to-one system, power enhancement can be achieved by increasing the receiving terminal voltage.

Table II compares the transmission characteristics of several WPT systems released in recent years. It can be seen that the PT-WPT system with double transmission coil structure proposed in this article has competitive advantages in terms of transmission power and resistance to spatial position changes, and does not increase the number of OA. The transmission distance is affected by the size of the coil, as the size of the coil used in this system is relatively small, resulting in a lower transmission distance. Therefore, we will not make too many comparisons here.

Table II: Comparison of Existing WPT Systems

| Existing references | The system architecture | The maximum transfer distance | Maximum output power | Power supply |
|---|---|---|---|---|
| **2017 [12]** | PP:one-to-one | 70 cm | 19.7 mW | OA |
| **2018 [13]** | SS:one-to-one | 47 cm | 2.4 mW | OA |
| **2024 [14]** | PP:one-to-one | 23 cm | 70 mW | OA |
| **2025 [15]** | SS:one-to-one | 32 cm | 30 W | Inverter |
| **2024 [16]** | PP:one-to-one | 23 cm | 70 mW | OA |
| **2024 [18]** | PP:one-to-one | 10 cm | 8 mW | OA |
| **2023 [26]** | SS:two-to-one | 17*10 cm(YZ) | 28 W | Inverter |
| **This paper** | PP:two-to-one | 13*17*14 cm(xyz)<br>7*15*12 cm(xyz) | 510 mW | OA |

## V. CONCLUSION

To enhance the misalignment tolerance capability and output power of the WPT system under spatial position variations, a novel OA negative-resistance structure was developed, accompanied by the proposal of a double-transmitting and single-receiving coil configuration. Both systems feature a circuit structure at the transmitter and receiver ends consisting of capacitors and inductors connected in parallel. Under PT symmetry two-to-one transmission conditions, the system achieves a maximum output power of 510 mW. While meeting the critical coupling coefficient requirement, the system maintains normal operation across a load resistance range from tens of ohms to tens of kilohms, significantly expanding the operational load range compared to conventional structures. Particularly in the PT symmetry two-to-one system, the large-sized receiving coil B enables stable power transfer with a receiving voltage 1.95 times the transmitting voltage within a horizontal (XY) plane area of approximately 13 cm × 17 cm and a vertical (Z) displacement of about 14 cm. Similarly, the small-sized receiving coil A achieves stable voltage output—also at 1.37 times the transmitting voltage over a horizontal (XY) area of 7 cm × 15 cm and a vertical (Z) range of 12 cm. The voltage fluctuation rate in both receiving systems remains below 3.4%, the actual transmission efficiency of both systems remains above 94.4%. Furthermore, compared to a one-to-one system under identical conditions, the two-to-one systems A and B exhibit increases in load voltage to 136% and 177%, and in output power to

<

185% and 313%, respectively.